\documentclass[aps,prl,twocolumn,amsmath,amssymb,showpacs,superscriptaddress,notitlepage,longbibliography]{revtex4-1}
\usepackage[colorlinks=true,linkcolor=blue,anchorcolor=red,citecolor=blue, urlcolor=blue]{hyperref}
\usepackage{bm}
\usepackage{graphicx}
\usepackage{color}

\begin{document}

\title{Disorder-induced nonlinear Hall effect with time-reversal symmetry}

\author{Z. Z. Du}
\affiliation{Shenzhen Institute for Quantum Science and Engineering and Department of Physics, Southern University of Science and Technology, Shenzhen 518055, China}
\affiliation{Shenzhen Key Laboratory of Quantum Science and Engineering, Shenzhen 518055, China}
\affiliation{Peng Cheng Laboratory, Shenzhen 518055, China}
\affiliation{School of Physics, Southeast University, Nanjing 211189, China}

\author{C. M. Wang}
\affiliation{Department of Physics, Shanghai Normal University, Shanghai 200234, China}
\affiliation{Shenzhen Institute for Quantum Science and Engineering and Department of Physics, Southern University of Science and Technology, Shenzhen 518055, China}
\affiliation{Shenzhen Key Laboratory of Quantum Science and Engineering, Shenzhen 518055, China}

\author{Shuai Li}
\affiliation{Shenzhen Institute for Quantum Science and Engineering and Department of Physics, Southern University of Science and Technology, Shenzhen 518055, China}
\affiliation{Shenzhen Key Laboratory of Quantum Science and Engineering, Shenzhen 518055, China}

\author{Hai-Zhou Lu}
\email{Corresponding author: luhaizhou@gmail.com}
\affiliation{Shenzhen Institute for Quantum Science and Engineering and Department of Physics, Southern University of Science and Technology, Shenzhen 518055, China}
\affiliation{Shenzhen Key Laboratory of Quantum Science and Engineering, Shenzhen 518055, China}
\affiliation{Peng Cheng Laboratory, Shenzhen 518055, China}

\author{X. C. Xie}
\affiliation{International Center for Quantum Materials, School of Physics, Peking University, Beijing 100871, China}
\affiliation{Beijing Academy of Quantum Information Sciences, Beijing 100193,China}
\affiliation{CAS Center for Excellence in Topological Quantum Computation, University of Chinese Academy of Sciences, Beijing 100190, China}

\date{\today }

\begin{abstract}
The nonlinear Hall effect has opened the door towards deeper understanding of topological states of matter. It can be observed as the double-frequency Hall voltage response to an ac longitudinal current in the presence of time-reversal symmetry. Disorder plays indispensable roles in various linear Hall effects, such as the localization in the quantized Hall effects and the extrinsic mechanisms of the anomalous, spin, and valley Hall effects. Unlike in the linear Hall effects, disorder enters the nonlinear Hall effect even in the leading order. However, the disorder-induced contribution to the nonlinear Hall effect has not been addressed.
Here, we derive the formulas of the nonlinear Hall conductivity in the presence of disorder scattering.
We apply the formulas to calculate the nonlinear Hall response of the tilted 2D Dirac model, which is the symmetry-allowed minimal model for the nonlinear Hall effect and can serve as a building block in realistic band structures.
More importantly, we construct the general scaling law of the nonlinear Hall effect, which may help in experiments to distinguish disorder-induced contributions to the nonlinear Hall effect. This work will be instructive for exploring unconventional responses upon breaking discrete or crystal symmetries in emergent physical systems and materials.
\end{abstract}

\maketitle

\begin{figure}[htpb]
\centering
  \includegraphics[width=0.48\textwidth]{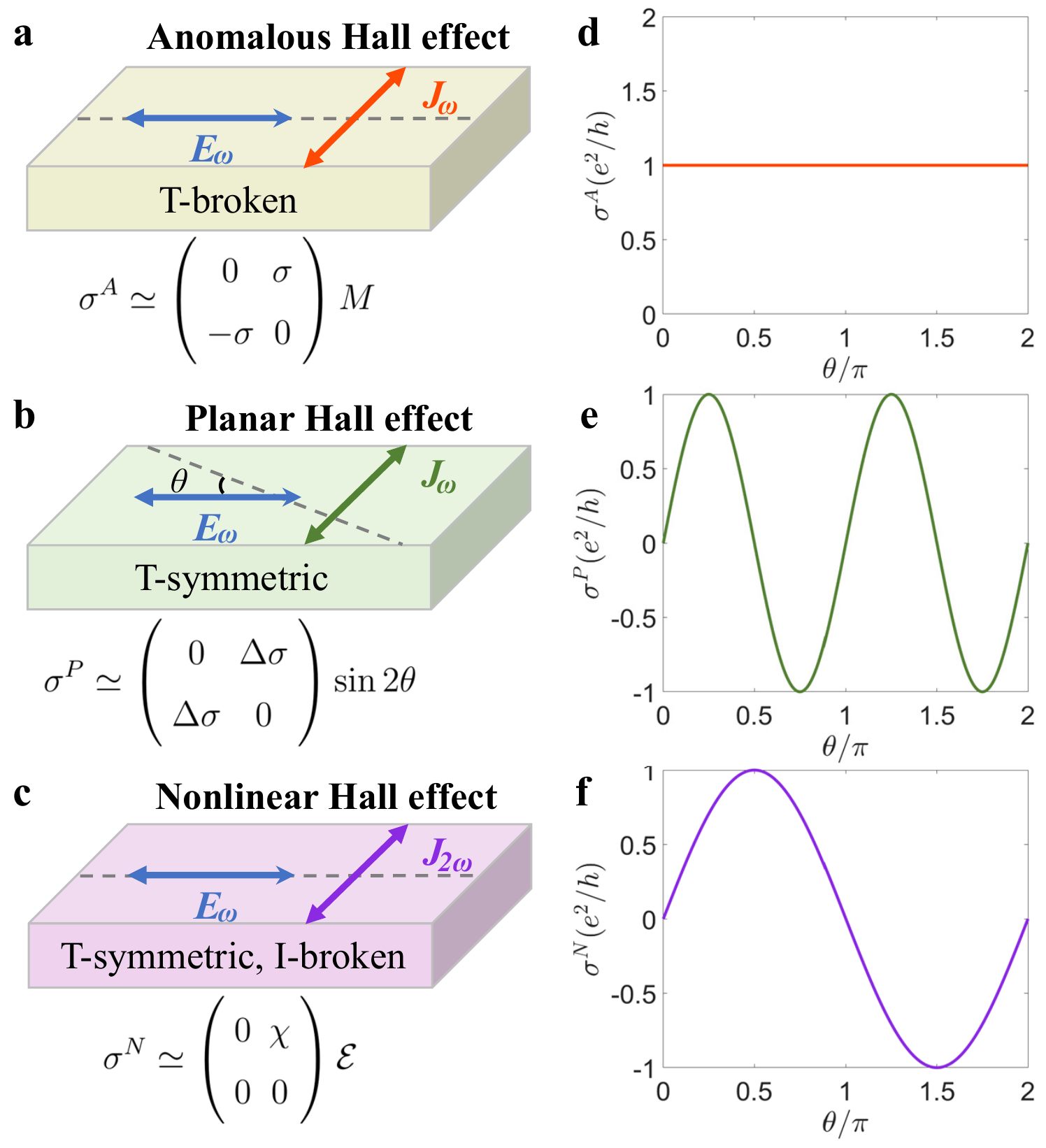}\\
  \caption{
  {\bf Comparison of the linear and nonlinear Hall effects in absence of the magnetic field.}
  Experimental setups and time-reversal symmetry of the anomalous (\textbf{a}), planar (\textbf{b}), and nonlinear Hall effects (\textbf{c}).
  $\sigma^A$ is the anomalous Hall conductivity, which is always anti-symmetric \cite{Nagaosa10rmp}. $M$ represents the magnetization. $\sigma^P$ is the planar Hall conductivity. $\Delta\sigma\equiv\sigma_{\parallel}-\sigma_{\perp}$, where $\sigma_{\parallel}$ and $\sigma_{\perp}$ refer to the longitudinal conductivities along the two principal axes. $\theta$ is the angle between the driving current and the $||$ principal axis (the dashed lines).
  $\sigma^N$ is the nonlinear Hall conductivity, which is proportional to the magnitude of the driving electric field $\mathcal{E}$. The element of the nonlinear Hall response tensor $\chi$ is due to inversion symmetry breaking along the dashed line.
  (\textbf{d}-\textbf{f}) Angular dependence can be used to distinguish the anomalous, planar, and nonlinear Hall effects. }\label{Fig:Hall}
\end{figure}

The Hall effects refer to a transverse voltage in response to a current applied in a sample of metal or semiconductor. The family of the classical and quantized Hall effects is one of the mainstreams of modern condensed matter physics, leading to the full spectrum of the search on the topological states of matter and many practical applications \cite{Klitzing80prl,QHE2012}.
All previous Hall effects are in the linear-response regime, that is, the transverse voltage is linearly proportional to the driving current, and a measurable Hall voltage requires that time-reversal symmetry is broken by magnetic fields or magnetism \cite{Klitzing80prl,QHE2012,Nagaosa10rmp,yasuda2016NP}.
The recently discovered nonlinear Hall effect \cite{Sodemann15prl,Low15PRB,Facio18PRL,You18PRB,Ma18nat,Kang18NM,Du18prl} does not need time-reversal symmetry breaking but inversion symmetry breaking, significantly different from the known linear Hall effects (Fig.~\ref{Fig:Hall}).
The linear Hall effects can be understood in terms of the Berry curvature \cite{Thouless82prl}, which describes bending of a parameter space (real space, momentum space, any vector fields) \cite{Xiao10rmp}. This geometric description is of the same significance as the curved spacetime in the general theory of relativity. The nonlinear Hall effect depends on the higher-order properties of the Berry curvature, thus not only can bring our knowledge to the next level but also may help device applications. More importantly, by adjusting the measurements to the nonlinear regime, a new territory is presented, in which unconventional responses upon breaking discrete and crystal symmetries can be studied in a great number of emergent materials.

The disorder effects have been a large part of the research on the linear Hall effects, such as the localization in the quantized Hall effects \cite{Prange90book,Chang13sci}, the extrinsic mechanisms of the anomalous \cite{Nagaosa10rmp}, spin \cite{Sinova15rmp}, and valley \cite{Mak14sci} Hall effect, etc. The debate on the origin of the anomalous Hall effect lasted for one century, until recently the mechanisms are summarized in terms of intrinsic (disorder-free) and extrinsic (disorder-induced) contributions \cite{Nagaosa10rmp}. The quantitative agreement between theories and experiments shows that the disorder-induced contribution are comparably important \cite{tian2009PRL,hou2015PRL}. In the nonlinear Hall effect, disorder is more important, because the effect always requires that the Fermi energy crosses an energy band. On the Fermi surface, the disorder scattering is inevitable and enters the nonlinear Hall effect even in the leading order.
This is quite different from the disorder-free leading order in the linear Hall effects. How disorder contribute to the nonlinear Hall signal remains unknown and is the focus of the investigations at this stage.

In this work, we use the Boltzmann equation formalism to derive the formulas of the nonlinear Hall conductivity in the presence of disorder scattering. The formulas can be applied to different
models to calculate the nonlinear Hall responses.
We apply the formula to the 2D tilted massive Dirac model.
The model is a symmetry-allowed minimal model for the
nonlinear Hall effect and can be used to understand the nonlinear Hall signals in realistic band structures \cite{Du18prl}.
Depending on roles of disorder scattering, we follow the convention to classify the nonlinear Hall conductivity into the ``intrinsic", side-jump, and skew-scattering contributions. The latter two are new findings to the framework of the nonlinear Hall effect and comparably important. The competition between the three contributions can induce a sign change in the nonlinear Hall signal.
More importantly, we present the scaling laws of the nonlinear Hall effect, which help to identify distinct contributions and explain the temperature and thickness dependence in the experiments.

\begin{table*}[htb]
\centering
\caption{
\bf{Formulas of the anomalous and nonlinear Hall responses in the $\omega\tau\ll1$ limit.}}\label{Tab:Comparison}
\begin{ruledtabular}
\begin{tabular}{ccc}
& Anomalous Hall response ($e^{2}/\hbar$) & Nonlinear Hall response ($e^{3}/2\hbar^{2}$) \\
\vspace{2mm}
Time-reversal symmetry & Broken  & Preserved \\
\vspace{2mm}
Intrinsic & $\sigma^{in}_{ab}=-\sum_{l}\varepsilon^{abc}\Omega^{c}_{l}f^{(0)}_{l}$
          & $\chi^{in}_{abc}=-\sum_{l}\varepsilon^{acd}\Omega^{d}_{l}g^{b}_{l}$ \\
Side-jump (velocity) & $\sigma^{sj,1}_{ab}=-\sum_{l}v_{a}^{sj}g^b_l$
          & $\chi^{sj,1}_{abc}=-\sum_{l}\tau_{l}v_{a}^{sj}\partial_cg^{b}_{l}$ \\
\vspace{2mm}
Side-jump (distribution) & $\sigma^{sj,2}_{ab}=\sum_{l}v_{b}^{sj}g^a_l$
          & $\chi^{sj,2}_{abc}=-\hbar\sum_{l}\tau_{l}\{[\partial_{a}(\tau_{l}v^{sj}_{c})
+\tilde{\mathcal{M}}^{ac}_{l}]v^{b}_{l}+\partial_{c}(\tau_{l}v^{a}_{l})v^{sj}_{b}\}
\frac{\partial f^{(0)}_{l}}{\partial\varepsilon_{l}}$  \\
\vspace{2mm}
 Intrinsic skew-scattering
          & $\sigma^{sk,1}_{ab}=-\sum_{ll'}\varpi^{g}_{ll'}\mathcal{U}^{a}_{ll'}g^{b}_{l}$
          & $\chi^{sk,1}_{abc}=\sum_{ll'}\varpi^{g}_{ll'}(\tilde{\mathcal{U}}^{ca}_{ll'}-\tau_{l}\mathcal{U}^{a}_{ll'}\partial_{c})g^{b}_{l}$  \\
Extrinsic skew-scattering
          & $\sigma^{sk,2}_{ab}=-\sum_{ll'}\varpi^{ng}_{ll'}\mathcal{U}^{a}_{ll'}g^{b}_{l}$
          & $\chi^{sk,2}_{abc}=\sum_{ll'}\varpi^{ng}_{ll'}(\tilde{\mathcal{U}}^{ca}_{ll'}-\tau_{l}\mathcal{U}^{a}_{ll'}\partial_{c})g^{b}_{l}$ \\
\end{tabular}
\end{ruledtabular}
\begin{flushleft}
We refer to the leading-order of the nonlinear Hall conductivity as the intrinsic contribution, but it depends on the disorder scattering, quite different from the disorder-free intrinsic Hall conductivity. The side-jump and skew-scattering contributions are due to the coordinates shift and antisymmetric scattering, respectively.
Here $\varepsilon^{abc}$ is the anti-symmetric tensor, we define $\partial_a\equiv\partial/\partial k_a$, $\partial'_a\equiv\partial/\partial k'_a$ and $g^{a}_{l}\equiv\tau_{l}\partial_af^{(0)}_{l}$.
The Berry curvature \cite{Xiao10rmp,Nagaosa10rmp} $\Omega^{a}_{l}=-2\varepsilon^{abc}\sum_{l'\neq l}\mathrm{Im}\langle l|\partial_{b}\hat{\mathcal{H}}|l'\rangle\langle l'|\partial_{c}\hat{\mathcal{H}}|l\rangle$/$(\varepsilon_{l}-\varepsilon_{l'})^{2}$, where $|l\rangle$ is the eigen vector.
The side-jump velocity $v_{a}^{sj}=\sum_{l'}\varpi^{sy}_{ll'}\delta r^{a}_{l'l}$ and $\tilde{\mathcal{M}}^{ab}_{l}\equiv\sum_{l'}\big(\tilde{M}^{ac}_{ll'}-\tilde{M}^{ac}_{l'l}\big)\delta(\varepsilon_{l}-\varepsilon_{l'})$, where $\varpi^{sy}_{ll'}$ is the symmetric scattering rate, the coordinates shift \cite{sinitsyn2007JPCM} $\delta r^a_{ll'}= i\langle l |\partial_a |l\rangle-i\langle l'|\partial'_{a}|l'\rangle-(\partial_a+\partial'_a)\arg(V_{ll'})$ with $V_{ll'}\equiv\langle l|\hat{V}_{imp}|l'\rangle$ and $\tilde{M}^{ab}_{ll'}\equiv(2\pi/\hbar)\partial_{a}(\tau_{l}|T_{ll'}|^{2}\delta r^{b}_{ll'})$.
$\varpi^{g}_{ll'}$ and $\varpi^{ng}_{ll'}$ refer to the Gaussian and non-Gaussian antisymmetric scattering rate, and we define that $\mathcal{U}^{a}_{ll'}\equiv\tau_{l}v^{a}_{l}-\tau_{l'}v^{a}_{l'}$ and $\tilde{\mathcal{U}}^{ab}_{ll'}\equiv\tau_{l}\partial_{a}(\tau_{l}v^{b}_{l})-\tau_{l'}\partial'_{a}(\tau_{l'}v^{b}_{l'})$.
\end{flushleft}
\end{table*}

\section{Formulas for Disorder-induced nonlinear Hall conductivity}

The nonlinear Hall effect is measured as zero- and double-frequency transverse electric currents driven by a low-frequency $ac$ longitudinal electric field $J_a $ in the absence of magnetic field \cite{Sodemann15prl,zhang2018TDM}, where the the $ac$ electric field $E_b(t)=\mathrm{Re}\{\mathcal{E}_be^{i\omega t}\}$ with the amplitude vector $\mathcal{E}_b$ and frequency $\omega$. The current up to the second-order of the $ac$ electric field can be found as $J_{a}=\mathrm{Re}\big\{J^{(0)}_{a}+J^{(1)}_{a}e^{i\omega t}+J^{(2)}_{a}e^{i2\omega t}\big\}$, with
\begin{eqnarray}
J^{(0)}_{a}=\xi_{abc}\mathcal{E}_{b}\mathcal{E}^{*}_{c}, \ \  J^{(1)}_{a}=\sigma_{ab}\mathcal{E}_b, \ \
J^{(2)}_{a}=\chi _{abc}\mathcal{E}_{b}\mathcal{E}_{c},
\end{eqnarray}
respectively, where $\{a,b,c\}\in \{ x,y,z\}$.
Table \ref{Tab:Comparison} summarizes our main results
for the anomalous Hall response tensor $\sigma_{ab}$ and the double-frequency nonlinear Hall response tensor $\chi_{abc}$ (see Methods).
We have assumed that $\omega \tau \ll 1$, because $\omega$ is about tens of Hertz and $\tau$ is about picoseconds in experiments.
This low-frequency limit is one of the differences from the nonlinear optics.
The disorder-induced zero-frequency response $\xi_{abc}$ is identical with the double-frequency response $\chi_{abc}$ in the $\omega\tau \ll 1$ limit.
Away from the $\omega\tau\ll 1$ limit, the double- and zero-frequency nonlinear Hall conductivities have different frequency dependence, thus are different in general.
In Sec.~SIII of \cite{Supp} we list the $\omega$-dependent full expressions with and without time-reversal symmetry, which would be helpful for understanding the recently proposed high-frequency rectification \cite{Isobe18arXiv} and gyrotropic Hall effects \cite{Konig18arXiv}.
According to how disorder works, the formulas are classified in terms of intrinsic ($in$), side-jump ($sj$), skew-scattering ($sk$) contributions. The formulas in Table \ref{Tab:Comparison} can be applied to different models to calculate the nonlinear Hall responses.

\section{Disorder-induced Nonlinear Hall conductivity of 2D Tilted massive Dirac model}

Now we apply Table \ref{Tab:Comparison} to calculate the nonlinear Hall conductivity in the presence of disorder scattering,
for the tilted 2D massive Dirac model (see Methods).
The model gives the symmetry-allowed minimal description of the nonlinear Hall effect and can serve as a building block in realistic band structures \cite{Du18prl}
\begin{eqnarray}\label{Eq:TiltedDirac}
\hat{\mathcal{H}}=tk_x+v(k_x\sigma_{x}+k_y\sigma_{y})+m\sigma_{z},
\end{eqnarray}
where $(k_{x}, k_{y})$ are the wave vectors, $(\sigma_{x}, \sigma_{y}, \sigma_{z})$ are the Pauli matrices, $m$ is the mass term, and $t$ tilts the Dirac cone along the $x$ direction. The time reversal of the model contributes equally to the Berry dipole, so it is enough to study this model only. For the disorder part, we assume a $\delta$-correlated spin independent random potential $\hat{V}_{imp}(\mathbf{r})=\sum_{i}V_{i}\delta(\mathbf{r}-\mathbf{R}_{i})$ with both Gaussian $\langle V^{2}_{i}\rangle_{dis}=V^{2}_{0}$ and non-Gaussian correlations $\langle V^{3}_{i}\rangle_{dis}=V^{3}_{1}$.

\begin{figure}[htpb]
\centering
  \includegraphics[width=0.4\textwidth]{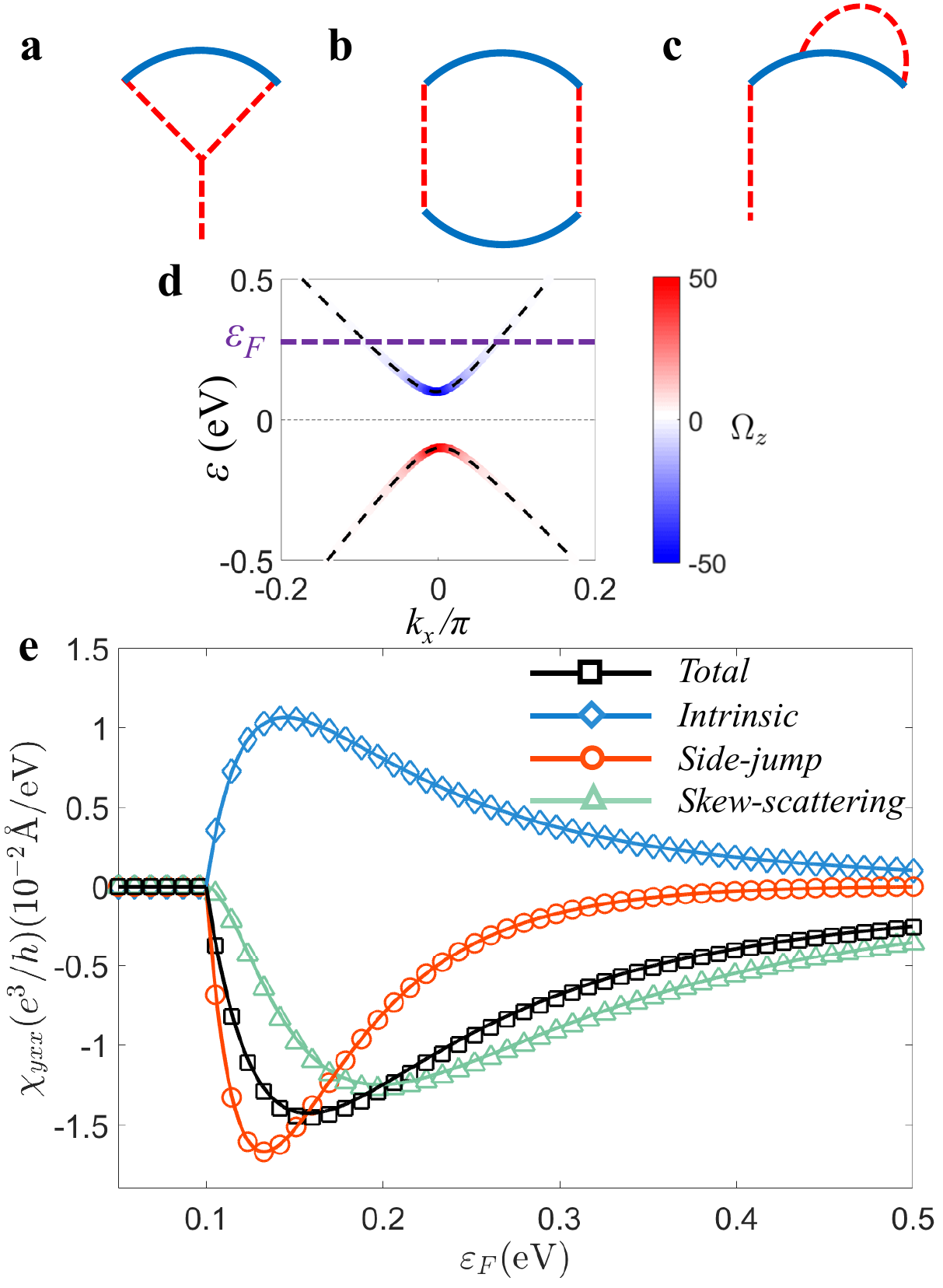}\\
  \caption{
  {\bf Nonlinear Hall response of 2D tilted massive Dirac model.}
  Terms contributing to the antisymmetric part of the scattering rate $\varpi^{(3)}_{ll'}$ (\textbf{a}) and $\varpi^{(4)}_{ll'}$ (\textbf{b} and \textbf{c}).
  \textbf{d} The band structure with the intensity plot of the Berry curvature $\Omega_z$. \textbf{e} The intrinsic, side-jump, skew-scattering and total contributions to nonlinear Hall conductivity $\chi_{yxx}$ of the 2D tilted massive Dirac model at zero temperature with a constant relaxation time $\tau$.
  The markers are the numerical results and the solid lines are analytic results up to leading $t$.
  Parameters are chosen as $t=0.1~\mathrm{eV\cdot\AA}$, $v=1~\mathrm{eV\cdot\AA}$, $m=0.1~\mathrm{eV}$, $n_{i}V^{2}_{0}=10^2~\mathrm{eV}^2\cdot\mathrm{\AA}^2$ and $n_{i}V^{3}_{1}=10^4~\mathrm{eV}^3\cdot\mathrm{\AA}^4$.}\label{Fig:Conductivity}
\end{figure}

To have analytic expressions with intuitive insight, we assume that $t\ll v$ and the scattering time is $k$-independent, i.e., $1/\tau=n_{i}V^{2}_{0}(\varepsilon^{2}_{F}+3m^2)/(4\hbar v^{2}\varepsilon_{F})$ (see details in Sec.~SIV of \cite{Supp}).
As functions of the Fermi energy $\varepsilon_{F}$, we obtain the analytic expressions for the intrinsic
\begin{eqnarray}\label{Eq:In}
\chi^{in}_{yxx}
=\frac{e^3}{h }\frac{t m}{n_iV_0^2}
\frac{3v^{2}(\varepsilon^{2}_{F}-m^{2})}{2\varepsilon^{3}_{F}(\varepsilon^{2}_{F}+3m^2)},
\end{eqnarray}
side-jump
\begin{eqnarray}
\chi^{sj}_{yxx}
&=&\frac{e^3}{h }\frac{t m}{n_iV_0^2}
\frac{v^{2}(\varepsilon^{2}_{F}-m^2)(\varepsilon^{2}_{F}-25m^2)}{2\varepsilon^{3}_{F}(\varepsilon^{2}_{F}+3m^2)^{2}},
\end{eqnarray}
and skew-scattering response functions
\begin{eqnarray}
\chi^{sk,1}_{yxx}
&=&-\frac{e^3}{h }\frac{t m}{n_iV_0^2}
\frac{v^{2}(\varepsilon^{2}_{F}-m^2)^{2}(13\varepsilon^{2}_{F}+77m^2)}
{4\varepsilon^{3}_{F}(\varepsilon^{2}_{F}+3m^2)^{3}},
\nonumber\\
\chi^{sk,2}_{yxx}
&=&-\frac{e^3}{h }\frac{t m }{n_i^2 V_0^6/V_1^3 } \frac{v^{2}(\varepsilon^{2}_{F}-m^2)^{2}(5\varepsilon^{2}_{F}+9m^2)}
{\varepsilon^{2}_{F}(\varepsilon^{2}_{F}+3m^2)^{3}}
\end{eqnarray}
up to the linear order in $t$. $n_i$ is the impurity density. $\chi_{xyy}=0$ for each contribution, as required by mirror reflection symmetry $k_{y}\leftrightarrow-k_{y}$.
According the above analytic expressions, the side-jump ($\chi^{sj}$) and intrinsic skew-scattering ($\chi^{sk,1}$) contributions are of the same order with the intrinsic one ($\chi^{in}$).
The extrinsic skew-scattering ($\chi^{sk,2}$) contribution is controlled by the relative scattering strength of the non-Gaussian scattering $V_1^3$.
The factor $\varepsilon^{2}_{F}-m^2$ in all the contributions secures that the nonlinear Hall conductivity vanishes at the band edge.
It is interesting to note that the side-jump contribution dominates near the bottom of the band, which is consistent with the result of a recent work \cite{Nandy19arXiv}.
At higher $\varepsilon_F$, the skew-scattering becomes the strongest contribution, which is similar to the behaviors in the anomalous Hall effect. All contributions vanish as $\varepsilon_F\rightarrow \infty$.  These behaviors can be seen in Fig.~\ref{Fig:Conductivity} \textbf{e}.

The zero-frequency nonlinear Hall response was not addressed experimentally.
In the $\omega\tau\ll 1$ limit, $\xi_{yxx}=\chi_{yxx}$. According to symmetry, $\xi_{xyy}=0$.
In the $dc$ limit ($\omega=0$), the electric field becomes time independent $E_{a}(t)=\mathcal{E}_{a}$,
and the nonlinear Hall response becomes a $dc$ current $J_{a}=(\xi_{abc}+\chi_{abc})\mathcal{E}_{b}\mathcal{E}_{c}=2\chi_{abc}\mathcal{E}_{b}\mathcal{E}_{c}$, which means that for the Dirac model tilted along the $x$ direction [Eq.~(\ref{Eq:TiltedDirac})], an $x$-direction
electric field can generate a \emph{dc} nonlinear Hall current along the $y$ direction. As a result, the measured Hall conductivity will be proportional to the electric field
\begin{equation}
\sigma^{N}_{yx}=2\chi_{yxx}\mathcal{E}_{x}.
\end{equation}
In contrast, if the electric field is along the $y$ direction, there is no such a Hall signal because
$\chi_{xyy}=0$, as required by the $y$-direction mirror reflection symmetry.
This indicates that the $dc$ nonlinear Hall signal $\sigma^{N}_{xy}$ has one-fold angular dependence. This $dc$ Hall signal can exist in the presence of time-reversal symmetry, which has been observed in the nonmagnetic
Weyl-Kondo semimetal Ce$_3$Bi$_4$Pd$_3$ \cite{Dzsaber18arXiv}.

\section{Scaling law of nonlinear Hall effect}

It is of fundamental importance to distinguish the different contributions to the nonlinear Hall signal in experiments.
For the anomalous Hall effect,
distinguishing different contributions is based on the scaling law of the transverse Hall signal to the longitudinal signal \cite{Nagaosa10rmp,tian2009PRL,hou2015PRL,yue2016JPSJ}.
For the nonlinear Hall effect, a scaling law can be constructed as well.
We adopt the quantity $V^{N}_{y}/(V^{L}_{x})^{2}=\xi_{yxx}\rho_{xx}$ or $\chi_{yxx}\rho_{xx}$ as the experimental scaling variable \cite{Supp},
where $V^{N}_{y}$ and $V^{L}_{x}$ refer to the nonlinear Hall (zero- or double-frequency) and linear longitudinal voltage, respectively. To measure the nonzero $\chi_{yxx}$, the driving electric current is applied along the $x$ direction and the nonlinear Hall voltage is measured along the $y$ direction. An advantage of this variable is that the intrinsic and side-jump parts become disorder independent.

To account for multiple sources of scattering \cite{hou2015PRL,yue2016JPSJ},
we consider the scaling law of nonlinear Hall effect in a general manner.
For simplicity, we assume no correlation between different scattering sources, thus each source contributes to the total resistivity independently, as dictated by Matthiessen¡¯s rule $\rho_{xx}=\sum_{i}\rho_{i}$ \cite{Mahan1990}, where $\rho_{i}$ is the contribution of the $i$th type of disorder scattering
to the longitudinal resistivity.
According to Table~\ref{Tab:Comparison}, the general scaling law of the nonlinear Hall effect can be obtained as (see details in Sec.~SV of \cite{Supp})
\begin{equation}\label{Eq:Scaling}
\frac{V^{N}_{y}}{(V^{L}_{x})^{2}}=\mathcal{C}^{in}+\sum_{i}\mathcal{C}^{sj}_{i}\frac{\rho_{i}}{\rho_{xx}}
+\sum_{ij}\mathcal{C}^{sk,1}_{ij}\frac{\rho_{i}\rho_{j}}{\rho^{2}_{xx}}
+\sum_{i\in S}\mathcal{C}^{sk,2}_{i}\frac{\rho_{i}}{\rho^{2}_{xx}}.
\end{equation}
Here the disorder-independent coefficients are for the intrinsic ($\mathcal{C}^{in}$),
side-jump ($\mathcal{C}^{sj}_{i}$),
intrinsic skew-scattering ($\mathcal{C}^{sk,1}_{i}$), and extrinsic skew-scattering ($\mathcal{C}^{sk,2}_{ij}$) contributions, respectively.
$S$ stands for static disorder scattering sources \cite{Supp,hou2015PRL}.
To use Eq.~(\ref{Eq:Scaling}), one needs to specify scattering sources.
As an example, we consider two major scattering sources as those in \cite{hou2015PRL}, one static ($i=0$) and one dynamic ($i=1$), then the scaling law becomes
\begin{equation}
\frac{V^{N}_{y}}{(V^{L}_{x})^{2}}
=\frac{1}{\rho^{2}_{xx}}\Big(\mathcal{C}_{1}\rho_{xx0}+\mathcal{C}_{2}\rho^{2}_{xx0}+\mathcal{C}_{3}\rho_{xx0}\rho_{xxT}+\mathcal{C}_{4}\rho^{2}_{xxT}\Big),
\end{equation}
with four scaling parameters
\begin{eqnarray}\label{C1234}
\mathcal{C}_{1}&=&\mathcal{C}^{sk,2}, \ \ \mathcal{C}_{2}=\mathcal{C}^{in}+\mathcal{C}^{sj}_{0}+\mathcal{C}^{sk,1}_{00}, \nonumber\\ \mathcal{C}_{3}&=&2\mathcal{C}^{in}+\mathcal{C}^{sj}_{0}+\mathcal{C}^{sj}_{1}+\mathcal{C}^{sk,1}_{01} ,\nonumber\\ \mathcal{C}_{4}&=&\mathcal{C}^{in}+\mathcal{C}^{sj}_{1}+\mathcal{C}^{sk,1}_{11}.
\end{eqnarray}
$\mathcal{C}_{1,2,3,4}$ can be extracted from experiments \cite{tian2009PRL,hou2015PRL,yue2016JPSJ}.
Here $\rho_{xx0}$ is the residual resistivity due to static impurities
at zero temperature and $\rho_{xxT}\equiv\rho_{xx}-\rho_{xx0}$ is due to
dynamic disorders (e.g., phonons) at finite temperature.

\begin{figure}[htpb]
\centering
  \includegraphics[width=0.38\textwidth]{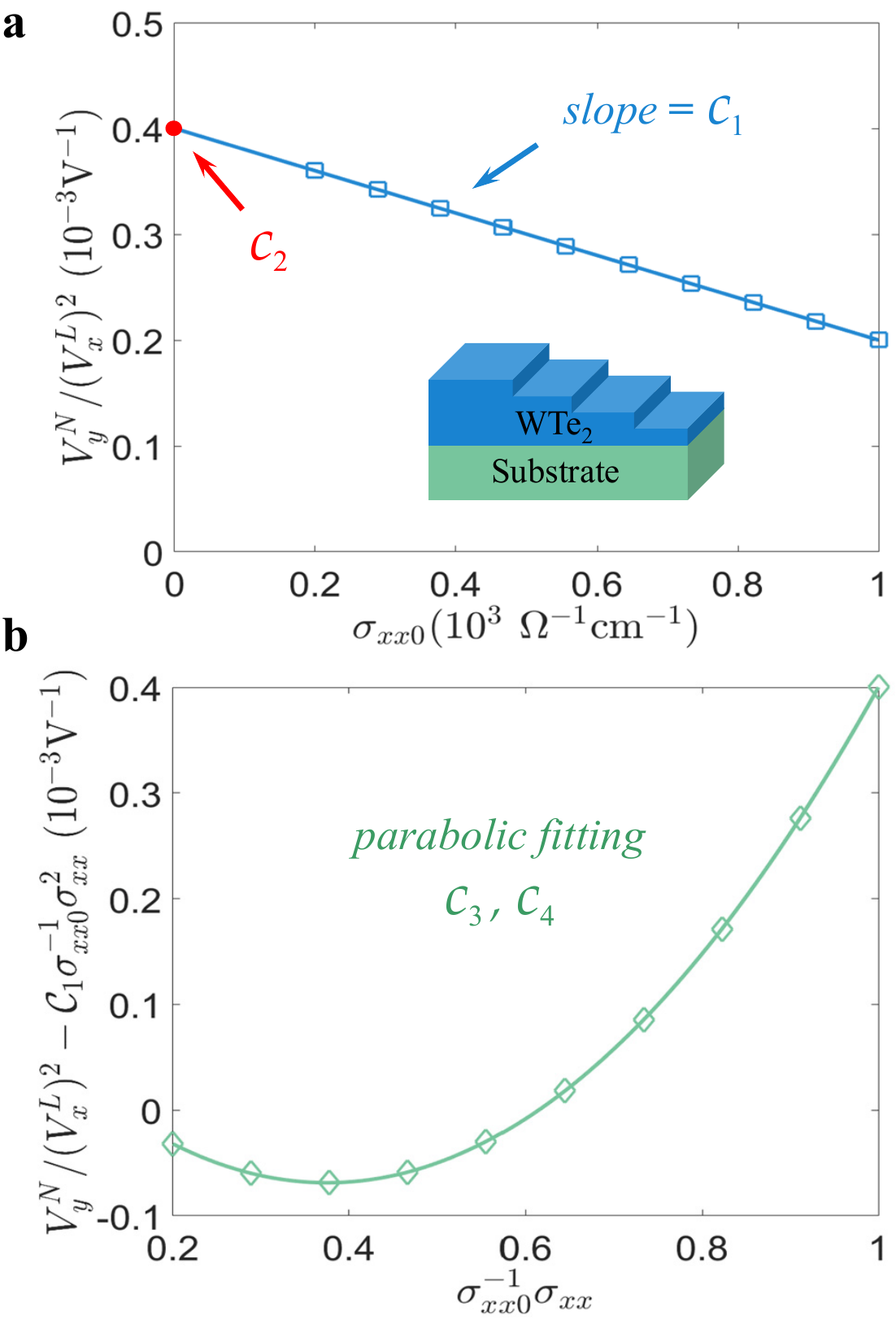}\\
  \caption{
 {\bf Scaling law of the nonlinear Hall effect.}
 \textbf{a} Step 1. At zero temperature, fitting $\mathcal{C}_1$ and $\mathcal{C}_2$ with the data of $V^{N}_{y}/(V^{L}_{x})^{2}$ and $\sigma_{xx0}$ for samples of different disorder strength (e.g., by changing the thickness \cite{tian2009PRL,hou2015PRL}).
 Insert is the schematic of the WTe$_2$ multi-step sample.
 \textbf{b} Step 2. At finite temperatures, for a given sample of known $\sigma_{xx0}$, fitting $\mathcal{C}_3$ and $\mathcal{C}_4$ with the data of $\sigma_{xx}$ at different temperatures. $\mathcal{C}_{1,2,3,4}$ can give most coefficients of physical meanings in Eq. (\ref{Eq:Scaling}).}\label{Fig:Scaling}
\end{figure}

In the zero-temperature limit ($T\rightarrow0$), we can approximate that $\rho_{xxT}\simeq0$ and $\rho_{xx}\simeq\rho_{xx0}=\sigma^{-1}_{xx0}$, then the scaling law becomes $V^{N}_{y}/(V^{L}_{x})^{2}\simeq\mathcal{C}_{1}\sigma_{xx0}+\mathcal{C}_{2}$, which indicates a linear scaling behaviour as shown in Fig.~\ref{Fig:Scaling} \textbf{a}.
Fitting the experimental data using this relation, the extrinsic skew-scattering coefficient $\mathcal{C}^{sk,2}$ can be experimentally extracted from the total nonlinear Hall conductivity (e.g., by using multi-step samples \cite{tian2009PRL,hou2015PRL,yue2016JPSJ}). Furthermore, at finite temperatures, it is more convenient to rewrite the scaling law into \begin{eqnarray}\label{Eq:DWL}
\frac{V^{N}_{y}}{(V^{L}_{x})^{2}}
-\mathcal{C}_{1}\sigma^{-1}_{xx0}\sigma^{2}_{xx}
&\simeq&
(\mathcal{C}_{2}+\mathcal{C}_{4}-\mathcal{C}_3)\sigma^{-2}_{xx0}\sigma^{2}_{xx}
\nonumber\\
&+&(\mathcal{C}_{3}-2\mathcal{C}_{4})\sigma^{-1}_{xx0}\sigma_{xx}+\mathcal{C}_{4}.
\end{eqnarray}
In this case, the proper scaling variable becomes $V^{N}_{y}/(V^{L}_{x})^{2}-\mathcal{C}_{1}\sigma^{-1}_{xx0}\sigma^{2}_{xx}$,
which is a parabolic function of $\sigma^{-1}_{xx0}\sigma_{xx}$ thus indicates a scaling behaviour shown in Fig.~\ref{Fig:Scaling} \textbf{b}.
By fitting the experimental data with the parabolic function, one can in principle extract the information of the rest scaling parameters, as shown in Fig.~\ref{Fig:Scaling}.
Equation~(\ref{Eq:DWL}) can be reorganized as $V^{N}_{y}/(V^{L}_{x})^{2} -\mathcal{C}_{1}\sigma^{-1}_{xx0}\sigma^{2}_{xx}\simeq (\mathcal{C}_{2}-\mathcal{C}_{4})\sigma^{-2}_{xx0}\sigma^{2}_{xx}
+(\mathcal{C}_{3}-2\mathcal{C}_{4})(\sigma^{-1}_{xx0}\sigma_{xx}-\sigma^{-2}_{xx0}\sigma^{2}_{xx})+\mathcal{C}_{4}$.
In the anomalous Hall effect, the second term on the right has been argued to be negligible in both the high-temperature limit ($\sigma_{xx0}\gg\sigma_{xx}$) and the low-temperature limit ($\sigma_{xx0}\simeq\sigma_{xx}$) \cite{tian2009PRL}. This linear scaling behaviour has been observed in thin films of WTe$_2$ \cite{Kang18NM}.
Nevertheless, Eq.~(\ref{Eq:DWL}) shows that the linear scaling behaviour with $\sigma^{2}_{xx}$  may become invalid in the high-conductivity regime \cite{hou2015PRL,yue2016JPSJ}. In the nonmagnetic Weyl-Kondo semimetal Ce$_3$Bi$_4$Pd$_3$, a linear scaling behavior of $\sigma^{N}_{xy}$ is observed as a function of $\sigma_{xx}$ \cite{Dzsaber18arXiv}.
The scaling law of voltages in Eq.~(\ref{Eq:DWL}) can be written into the scaling law of the nonlinear Hall conductivity
\begin{eqnarray}
\sigma^{N}_{xy}&\simeq&\mathcal{C}_{1}\sigma^{-1}_{xx0}\sigma^{3}_{xx}+(\mathcal{C}_{2}-\mathcal{C}_{3}+\mathcal{C}_{4})\sigma^{-2}_{xx0}\sigma^{3}_{xx}\nonumber\\
&&+(\mathcal{C}_{3}-2\mathcal{C}_{4})\sigma^{-1}_{xx0}\sigma^2_{xx}+\mathcal{C}_{4}\sigma_{xx},
\end{eqnarray}
for a fixed electric field. According to the conductivity scaling law, the observed linear behaviour in \cite{Dzsaber18arXiv} indicates the dominance of the scaling parameter $\mathcal{C}_{4}$. According to Eq. (\ref{C1234}), $\mathcal{C}_4$ is contributed mainly by the intrinsic mechanism and the dynamical scattering processes (e.g. Fig.~\ref{Fig:Conductivity} \textbf{b} and \textbf{c}).

\section{Methods}

\subsection{Boltzmann formulism in the nonlinear regime}

The nonlinear Hall effect is measured as zero- and double-frequency transverse electric currents driven by a low-frequency $ac$ longitudinal electric field $\mathbf{J}(\mathbf{E})=-e\sum_{l}\dot{\textbf{r}}_{l}\cdot n_{l}$ in the absence of a magnetic field, where the the $ac$ electric field $\mathbf{E}(t)=\mathrm{Re}\{\mathcal{E}e^{i\omega t}\}$ with the amplitude vector $\mathcal{E}$ and frequency $\omega$.
$-e$ is the electron charge and $l=(\eta,\mathbf{k})$ labels a state in band $\eta$ with wave vector $\mathbf{k}$.
The distribution function $n_{l}$ can be found from the standard Boltzmann equation \cite{Mahan1990}, which reads
\begin{eqnarray}
\frac{\partial f_{l}}{\partial t}+\dot{\textbf{k}}\cdot\frac{\partial f_{l}}{\partial\textbf{k}}
=\mathcal{I}_{el}\{f_{l}\}
\end{eqnarray}
in the spatially uniform case.
Here $\mathcal{I}_{el}\{f_{l}\}$ represents the elastic disorder scattering by static defects or impurities.
The elastic disorder scattering can be decomposed as the intrinsic, side-jump, and skew-scattering parts (see details in Sec. SI of \cite{Supp})
\begin{eqnarray}
\mathcal{I}_{el}\{f_{l}\}=\mathcal{I}^{in}_{el}\{f_{l}\}+\mathcal{I}^{sj}_{el}\{f_{l}\}+\mathcal{I}^{sk}_{el}\{f_{l}\}.
\end{eqnarray}
The intrinsic part is contributed by symmetric scatterings, in which incoming and outgoing states are reversible in a scattering event.
The side-jump part is resulting from the coordinates shift during scattering processes.
The skew-scattering part is contributed by anti-symmetric scatterings, in which exchanging the incoming and outgoing states yields a minus sign.
Specifically,
$\mathcal{I}^{in}_{el}\{f_{l}\}=-\sum_{l'}\varpi^{sy}_{ll'}(f_{l}-f_{l'})$,
$\mathcal{I}^{sj}_{el}\{f_{l}\}=-e\mathbf{E}\cdot\sum_{l'}\mathbf{O}_{ll'}(f_{l}-f_{l'})$,
$\mathcal{I}^{sk}_{el}\{f_{l}\}=-\sum_{l'}\varpi^{as}_{l'l}(f_{l}+f_{l'})$,
where $\varpi^{sy}_{ll'}$ and $\varpi^{as}_{ll'}$ represents the symmetric and antisymmetric parts of the scattering rate $\varpi_{ll'}=(2\pi/\hbar)|T_{ll'}|^{2}\delta(\varepsilon_{l}-\varepsilon_{l'})$ with $T_{ll'}$ representing the T-matrix \cite{Mahan1990,Supp}.
$\mathbf{O}_{ll'}\equiv(2\pi/\hbar)|T_{ll'}|^{2}\delta\mathbf{r}_{ll'}\frac{\partial}{\partial\varepsilon_{l}}\delta(\varepsilon_{l}-\varepsilon_{l'})$,
where the coordinates shift $\delta\mathbf{r}_{ll'}$ is defined in Table \ref{Tab:Comparison}.
The expression of $\dot{\mathbf{r}}_{l}$ and $\dot{\mathbf{k}}$ can be found from the semiclassical equations of motion \cite{sinitsyn2007JPCM,Xiao10rmp}
\begin{eqnarray}\label{Eq:EOM}
\dot{\textbf{r}}_{l}=\mathbf{v}_{l}-\dot{\textbf{k}}\times\mathbf{\Omega}_{l}
+\mathbf{v}^{sj}_{l}, \ \ \ \ \ \ \dot{\textbf{k}}=-\frac{e}{\hbar}\textbf{E},
\end{eqnarray}
where $\mathbf{v}_{l}=\partial\varepsilon_{l}/\hbar\partial\textbf{k}$ is the group velocity, $\mathbf{\Omega}_l$ is the Berry curvature \cite{Xiao10rmp}, and $\mathbf{v}^{sj}_{l}$ is the side-jump velocity \cite{sinitsyn2007JPCM} (see Table \ref{Tab:Comparison}).
To solve the Boltzmann equations up to the second order of $\mathbf{E}$, we adopt the relaxation time approximation \cite{Mahan1990} for the intrinsic scattering parts $\mathcal{I}^{in}_{el}\{f_{l}\}=(f^{(0)}_{l}-f_{l})/\tau_{l}$, where $f^{(0)}_{l}$ is the Fermi distribution function and $\tau_{l}$ represents the relaxation time.
Usually, in good metal regime, $\tau_{l}$ is treated as a constant that can be determined by experiments.
For systems with large anisotropy, $\tau_{l}$ can have a significant angular dependence  \cite{schliemann2003PRB,Xiao17PRB}.
With the above equations, the current up to the second-order responses to the $ac$ electric field can be obtained.

\subsection{Tilted 2D massive Dirac model with disorder}

We use the tilted 2D massive Dirac model to calculate the nonlinear Hall conductivity in Fig. \ref{Fig:Conductivity}.
\begin{eqnarray}\label{Eq:TiltedDirac}
\hat{\mathcal{H}}=tk_x+v(k_x\sigma_{x}+k_y\sigma_{y})+m\sigma_{z},
\end{eqnarray}
where $(k_{x}, k_{y})$ are the wave vectors, $(\sigma_{x}, \sigma_{y}, \sigma_{z})$ are the Pauli matrices, $m$ is the mass term, and $t$ tilts the Dirac cone along the $x$ direction.  The time reversal of the model contributes equally to the Berry dipole, so it is enough to study this model only. The model describes two energy bands (denoted as $\pm$) with the band dispersions $\varepsilon^{\pm}_{\mathbf{k}}=tk_{x}\pm[v^{2}k^{2}+m^{2}]^{1/2}$, where $k^{2}\equiv k^{2}_{x}+k^{2}_{y}$.
In the $x-y$ plane, the Berry curvature behaves like a pseudoscalar, with only the $z$ component $\Omega^z_{\pm \mathbf{k}}=\mp mv^{2}/[2(v^{2}k^{2}+m^{2})^{3/2}]$.

To consider the disorder effect, we expanded the scattering rate up to the fourth order in the disorder strength as $\varpi_{ll'}=\varpi^{(2)}_{ll'}+\varpi^{(3)}_{ll'}+\varpi^{(4)}_{ll'}$.
Here $\varpi^{(2)}_{ll'}$ is pure symmetric and of order $n_{i}V^{2}_{0}$ with $n_{i}$ refers to the concentration of disorder.
Fig.~\ref{Fig:Conductivity} \textbf{a} corresponds to the contribution to $\varpi^{(3)}_{ll'}$, which is non-Gaussian and of order $n_{i}V^{3}_{1}$.
Fig.~\ref{Fig:Conductivity} \textbf{b} and \textbf{c} correspond to $\varpi^{(4)}_{ll'}$ within non-crossing approximation, which is Gaussian and of order $n^{2}_{i}V^{4}_{0}$.
Thus, $\varpi^{(2)}_{ll'}$ is the leading symmetric contribution, $\varpi^{(3)}_{ll'}$ and $\varpi^{(4)}_{ll'}$ contain the leading non-Gaussian and Gaussian antisymmetric contribution to the scattering rate.
Considering all the leading contributions, we identify that $\varpi^{sy}_{ll'}=\varpi^{(2)}_{ll'}$ and $\varpi^{as}_{ll'}=\varpi^{(3a)}_{ll'}+\varpi^{(4a)}_{ll'}$, where $\varpi^{(3a)}_{ll'}$ and $\varpi^{(4a)}_{ll'}$ represent the antisymmetric parts of the third and fourth order scattering rate, respectively (see details in Sec.~SIV of \cite{Supp}).

\subsection{Code availability}
The code that is deemed central to the conclusions are available from the corresponding author upon reasonable request.

\subsection{Data availability}
The data that support the plots within this paper and other findings of this study are available from the corresponding author upon reasonable request.

\section{Acknowledgments}

We thank insightful discussions with Huimei Liu and Cong Xiao.
This work was supported by the Strategic Priority Research Program of Chinese Academy of Sciences (Grant No. XDB28000000),
the National Basic Research Program of China (Grant No. 2015CB921102), the National Key R \& D Program (Grant No. 2016YFA0301700),
the Guangdong Innovative and Entrepreneurial Research Team Program (Grant No. 2016ZT06D348),
the National Natural Science Foundation of China (Grants No. 11534001, No. 11474005, No. 11574127),
and the Science, Technology and Innovation Commission of Shenzhen Municipality (Grant No. ZDSYS20170303165926217, No. JCYJ20170412152620376).

\section{Author contributions}

All authors performed the calculations, discussed the results and prepared the manuscript.

\section{Competing financial interests}
The authors declare no competing financial interests.


%

\end{document}